%Paper: astro-ph/9302018
%From: BEST%IASSNS.BITNET@PUCC.PRINCETON.EDU
%Date: Thu, 25 Feb 93 11:22 EST

\input phyzzx
\titlepage
{\bf
\title{CONSTRAINTS ON THE COSMIC RAYS IN THE SMALL MAGELLANIC CLOUD}
\author{Arnon Dar\foot{Permanent Address: Technion, Israel Inst. of
Tech., Haifa 32000 ISRAEL}, Ari Laor and Abraham Loeb}
\address{School Of Natural Sciences, The Institute For Advanced Study,
Princeton, NJ 08540 USA}}
\abstract
We show that recent $\gamma$-ray observations of the Small Magellanic
Cloud with EGRET rule out a universal cosmic ray flux only at
energies below $\approx 10$ GeV, while the observed diffuse X-ray and
$\gamma$-ray background radiations  have already
ruled out, by more than three orders of magnitude, a universal
extragalactic cosmic ray flux
identical to that observed in the local solar neighborhood
at energies below $10^6$ GeV.

\submit{Physical Review Letters - Comments}
\endpage

In a recent letter\Ref\One{P. Sreekumar et al., Phys. Rev. Lett.
{\bf 70}, 127 (1993).} Sreekumar et al. reported an upper limit
(95\% confidence) of ${\rm F_\gamma(E>100~MeV)
< 0.5\times 10^{-7}~ photons~cm^{-2}s^{-1}} $
for the high energy $\gamma$-ray
flux from the Small Magellanic Cloud (SMC), while the expected flux that
was calculated
by Sreekumar and Fichtel\Ref\Two{P. Sreekumar and C.E. Fichtel,
Astron. Astrophys. {\bf 251}, 447, (1991).}, assuming that the cosmic
ray flux is universal and identical to that observed in the local
solar neighborhood, is ${\rm F_\gamma(E>100~MeV)\approx 2.4\times
10^{-7}}$ photons${\rm~cm^{-2}s^{-1}.}$ From this discrepancy
Sreekumar et al. concluded$^{^{[1]}}$ that the bulk of the
cosmic ray energy density is not universal
and must be galactic in origin.
\par
In this comment we show that the observations of Sreekumar
et al. rule out a universal cosmic ray flux only at
energies below $\approx 10$ GeV, while the observed diffuse X-ray and
$\gamma$-ray background radiations  have already
ruled out (by more than three orders of magnitude) a universal
extragalactic cosmic ray flux
identical to that observed in the local solar neighborhood
at energies below $10^6$ GeV.
\par
The production of high energy $\gamma$-rays by cosmic rays in the SMC
is dominated by the reaction $pp\rightarrow \pi^0X\rightarrow
2\gamma X$ of cosmic ray protons on interstellar hydrogen,
by bremsstrahlung from cosmic ray electrons in the interstellar gas
and by inverse Compton scattering of cosmic ray electrons on infrared,
visible and ultraviolet SMC photons. We limit our discussion to
these three processes.
\par
The product of the production cross section and $\pi^0$
multiplicity beyond threshold (at 279.6 MeV)
is a slowly increasing function of incident energy.
Folding this energy dependence with the differential cosmic
ray power law spectrum,
${\rm dF_p/dE\sim E^{-2.75}}$, one finds\Ref\Four{
S.A. Stephens and G.D. Badhwar, Astrophys. \& Space Sci. {\bf 76}, 213
(1981).} that the bulk $(>90\%)$ of the $\pi^0$'s and $\gamma$-rays with
E$_\gamma>100~$MeV are produced by cosmic ray protons of energy below
10 GeV.
\par
The  $\gamma$-ray flux from the SMC due to bremsstrahlung from
cosmic ray electrons in the interstellar gas is proportional to\Ref\Five{
F.W. Stecker, in "Origins of Cosmic Rays" (eds.  J.L. Osborne and A.W.
Wolfendale), p. 267 (Reidel, Dordrecht 1975).}
$ {\rm  F_e(>E_\gamma)/E_\gamma~,}$
where
${\rm F_e(>E_\gamma)}$ is the integral flux of electrons with
energy above $E_\gamma$.
If the high energy
cosmic ray electron flux is universal and identical to its
flux in the local solar neighborhood\Ref\Fe {K.R. Lang,
"Astrophysical Formulae" (Springer Verlag 1980), p. 471.},
${\rm dF_e/dE\approx 1.16\times 10^{-2}(E/GeV)^{-2.6} cm^{-2}
s^{-1}sr^{-1}GeV^{-1}~},$
then less than 2\% of the $\gamma$-rays with ${\rm E_\gamma>100~MeV}$ are due
to
bremsstrahlung from electrons with ${\rm E_e> 10~GeV~.}$
\par
The average energy of photons, whose initial average
energy is $\langle\epsilon\rangle$, that are Compton boosted
by the high energy cosmic ray electrons
(${\rm \gamma\equiv E_e/m_ec^2\gg 1)}$ is given by
${\rm \langle E_\gamma\rangle \approx (4/3)\gamma^2\langle\epsilon\rangle}~$
and their differential spectrum is proportional to
${\rm E_\gamma^{-(2.6+1)/2}=E_\gamma^{-1.8}}~.$ For the SMC emitted photons
$\langle \epsilon\rangle \approx $1 eV.
Thus, less than 3\% of the $\gamma$-rays
with $E_\gamma>$100 MeV from the SMC are expected to be
produced by cosmic ray electrons with energy larger than 10 GeV.
\par
Therefore, we conclude that the
EGRET observations of the SMC rule out a universal
cosmic ray flux only at energies below 10 GeV.
\par
But, if the high energy cosmic ray electrons were universal
with a density
${\rm dn_e/d\gamma \approx C\gamma^{-p}}~,$ they
would have produced diffuse X-ray and $\gamma$-ray cosmic backgrounds
by Compton boosting of microwave background photons
in the intergalactic space
to X-ray and $\gamma$-ray energies\Ref\FM{J.E. Felten and P. Morrison,
Ap. J. {\bf 146}, 686 (1966).}, with a
flux\Ref\zzz{G.B. Rybicki and A.P.
Lightman, "Radiative Processes In Astrophysics" (John Wiley 1979) p. 208.}
$${\rm {dF_\gamma \over dE}\approx { \sigma_{T}t_0C\over c h^3}
  {2^{p+1}3(p^2+4p+11)\over
   (p+3)^2(p+5)(p+1)}
   \Gamma \left({p+5}\over 2\right)\zeta \left({p+5\over 2}\right)
   (kT)^{(p+5)/2}E^{-(p+1)/2}}~, \eqno\eq $$
where ${\rm \sigma_T=6.5\times 10^{-25}cm^2}$
is the Thomson cross section,  t$_0$
is the age of the Universe, T$\approx$2.736K
is the temperature of the microwave photons
and $\zeta$ denotes the Riemann zeta function.
For cosmic ray electrons in the local solar neighborhood$^{^{[5]}}$
p$\approx 2.6$ and ${\rm C\approx 9\times 10^{-7} cm^{-3}}.$
If this flux was universal it would have produced diffuse X-ray
and $\gamma$-ray backgrounds with a flux
$${\rm {dF_\gamma\over dE}\approx 3.3\times 10^4
\left[{t_0\over 10~Gy}\right]
\left[{E\over keV}\right]^{-1.8}~cm^{-2}s^{-1}
sr^{-1}keV^{-1}}, \eqno\eq $$
which is larger by more than three orders of magnitude (!) than the observed
flux of the X-ray\Ref\xxx{R. Giacconi et al., Phys. Rev. Lett.
{\bf 9}, 439 (1962). For a recent review see A.C. Fabian
and X. Barcons, Ann. Rev. Astron. Astrophys. {\bf 30}, 429 (1992).}
and $\gamma$-ray\Ref\yyy{C.E. Fichtel et al., Ap. J. {\bf 222}, 833 (1978).}
background radiations.
These estimates do not change significantly when redshift effects
are taken into consideration.

\centerline{{\bf Acknowledgement}}
The authors would like to thank J.N. Bahcall for useful comments.

\endpage
\refout

\end